\documentclass[sigconf]{acmart}

\AtBeginDocument{%
  }

\usepackage{xspace}             
\usepackage{booktabs}           
\usepackage{multirow}           
\usepackage{lscape}             
\usepackage{longtable}          
\usepackage{tabularx}           
\usepackage{amsmath}            
\usepackage{wrapfig}
\usepackage{soul}

\setcopyright{acmlicensed}
\copyrightyear{2018}
\acmYear{2018}
\acmDOI{XXXXXXX.XXXXXXX}
\acmConference[EASE 2025]{The 29th International Conference on Evaluation and Assessment in Software Engineering}
\acmISBN{978-1-4503-XXXX-X/2018/06}

\begin{document}

\title{Mitigating Configuration Differences Between Development and Production Environments: A Catalog of Strategies}

\author{Marcos Felipe Carvalho Nazário}
\email{carvalhonazario@gmail.com}
\orcid{0009-0006-7889-6515}
\affiliation{%
  \institution{Federal University of Pará}
  \streetaddress{Augusto Corrêa Street}
  \city{Belém}
  \state{Pará}
  \country{Brazil}
}

\author{Rodrigo Bonifácio}
\email{rbonifacio@unb.br}
\orcid{0000-0002-2380-2829}
\affiliation{%
  \institution{University of Brasília}
  \streetaddress{Darcy Ribeiro University Campus}
  \city{Brasília}
  \country{Brazil}
}

\author{Gustavo Pinto}
\email{gpinto@ufpa.br}
\affiliation{%
  \institution{Federal University of Pará}
  \streetaddress{Augusto Corrêa Street}
  \city{Belém}
  \state{Pará}
  \country{Brazil}
}

\renewcommand{\shortauthors}{Nazário et al.}

\newcommand{\rb}[1]{\textcolor{red}{\textbf{[#1] - Rodrigo}}}
\newcommand{\gnote}[1]{\textcolor{red}{\textbf{[#1] - Gustavo}}}
\newcommand{\novo}[1]{\textcolor{blue}{#1}}

\newcommand{\totalParticipants}{17\xspace}
\newcommand{\totalStrategies}{eight\xspace}
\newcommand{\totalCodes}{14\xspace}

\begin{abstract}
\textbf{Context:} The Configuration Management of the development and production environments is an important aspect of IT operations. However, managing the configuration differences between these two environments can be challenging, leading to inconsistent behavior, unexpected errors, and increased downtime.
\textbf{Objective:} In this study, we sought to investigate the strategies software companies employ to mitigate the configuration differences between the development and production environments. Our goal is to provide a comprehensive understanding of these strategies used to contribute to reducing the risk of configuration-related issues.
\textbf{Method:} To achieve this goal, we interviewed \totalParticipants participants and leveraged the Thematic Analysis methodology to analyze the interview data. These participants shed some light on the current practices, processes, challenges, or issues they have encountered.
\textbf{Results:} Based on the interviews, we systematically formulated and structured a catalog of \totalStrategies strategies that explain how software producing companies mitigate these configuration differences. These strategies vary from 1) creating detailed configuration management plans, 2) using automation tools, and 3) developing processes to test and validate changes through containers and virtualization technologies.
\textbf{Conclusion:} By implementing these strategies, companies can improve their ability to respond quickly and effectively to changes in the production environment. In addition, they can also ensure compliance with industry standards and regulations.
\end{abstract}





\maketitle

\section{Introduction}\label{sec:Introduction}


DevOps is a cultural and practical approach that bridges development and operations, enabling faster and more reliable software delivery~\cite{DevOpsTh9,AdoptingDevOps2019}. To boost delivery speed and IT efficiency, companies increasingly adopt DevOps. A core goal is automating the delivery pipeline, including environment configuration, which demands a clear configuration management plan. Automation tools are essential to standardize these procedures in DevOps.

In DevOps pipelines, managing configuration differences between development and production systems is a recurrent challenge. These differences can lead to inconsistencies in behavior~\citep{shahri2007software}, unexpected errors~\citep{bartusevics2013methodology}, and increased downtime~\citep{couch2010troubleshooting}, all of which can negatively impact team productivity, system performance, and availability. Although the literature on DevOps is fertile, no research study consolidates and details the practices that development teams leverage to mitigate the configuration differences between production and development environments.
This paper introduces and details a systematically compiled catalog of \totalStrategies strategies that companies employ to mitigate configuration differences between development and production systems. The study identifies the steps teams take and their perceived efficacy by interviewing industry professionals, such as software engineers, IT managers, and technical executives. The advantages of automating configuration routines and the difficulties development teams experience when automating the configuration pipeline are also highlighted. By reducing configuration-related problems and guaranteeing more seamless system operations, the study aims to increase system performance and availability. 

Our work is empirically grounded and it main contribution is to confirm the software companies' strategies to mitigate the configuration differences between the development and production environments founded in our previous work~\cite{myGreyWork}. The second is a more in-depth understanding of how these strategies reduce the risk of configuration-related issues and guarantee software systems' smooth and reliable operation. The third is an insight into the practices, processes, challenges, and issues faced by IT experts and other stakeholders in the management and deployment of systems.

Although the paper frames the problem within the broader context of DevOps, the focus of participant selection and strategy analysis emphasizes the use of Infrastructure as Code (IaC) — a key enabler of DevOps goals such as automation, repeatability, and environmental consistency. It also emphasizes how critical it is to comprehend the strategies utilized by software development organizations to mitigate the risk involved with handling configuration discrepancies. It can substantially impact the area of IT operations and software development by offering helpful advice and insights into this crucial subject.

\noindent
\textbf{Artifacts availability.} The data analyzed in this work are publicly available~\cite{myCodes}.

\section{Background}


Modern software development depends on a complex set of tools, methods, and practices to ensure smooth deployment across environments. A major challenge is handling configuration differences between development and production~\cite{shahri2007software}, which often cause inconsistencies and failures. This section reviews Software Configuration Management (SCM), DevOps, and CI/CD—key concepts for addressing such issues.

\textbf{SCM.}
It focuses on controlling software modifications and maintaining consistency between environments, especially as software evolves. Unlike traditional configuration management for hardware~\cite{tichyConfiguration1995}, it addresses the variability of platforms, dependencies, and runtime services. The software industry has embraced procedures that bring the development and production environments closer together in an effort to reduce this complexity, which has given rise to the DevOps field~\cite{DevOpsTh9} to reduce this complexity, aligning development and operations by promoting consistent configurations through automation and CI/CD.


\textbf{DevOps.}
According to Ebert et al.~\cite{ebert2016devops,DevOpsTh9}, DevOps means a culture shift toward collaboration between development, quality assurance, and operations. It is about fast, flexible development and provisioning business processes. It efficiently integrates development, delivery, and operations, thus facilitating a lean, fluid connection of these traditionally separated silos. Using automated development, deployment, and infrastructure monitoring, it combines the development and operations worlds. This strategy aids in delivering value more quickly and constantly, minimizing issues brought on by team member misunderstandings, and speeding up the problem-solving process.


\textbf{Continuous Integration and Deployment.}
According to Shahin et al.~\cite{shahin2017continuous}, Continuous Integration (CI) and Continuous Deployment (CD) are called continuous practices that serve organizations to accelerate their development and delivery of software components without compromising quality. The CI is a widely established development practice in software development industry, in which members of a team integrate and merge development work (e.g., code) frequently, for example multiple times per day. The CD practice takes a step further and distributes the application to client or production environments automatically and continually. There is a robust debate in academic and industrial circles about defining and distinguishing between continuous deployment and continuous delivery~\cite{fitzgerald2017continuous}.

\section{Research Questions}

This research aims to understand how software companies mitigate development and production configuration. This embraces the differences in environment configuration, security, network routes, and further resources. Aiming for this goal, we focus on the following research questions:

\begin{itemize}
    \item [RQ1:] What actions are taken by the team to mitigate differences between the development and production environments?
    \item [RQ2:] Are these team actions sufficient to mitigate differences between the development and production environments?
    \item [RQ3:] What are the perceived benefits of automation in development and production environments?
    \item [RQ4:] What are the perceived difficulties in the automation process in development and production environments?
\end{itemize}

Answering \textbf{RQ1} is essential to understand the strategies used in industries to mitigate the differences between the development and production environments. In particular, knowing that the DevOps culture is an important tool to decrease the silos between the infrastructure and development teams, we sought to discuss this issue with professional DevOps engineers.
Answering \textbf{RQ2} is relevant to collect the perception of the interviewers if these strategies of the industries have been enough to mitigate these differences. Answering \textbf{RQ3} is pertinent to understand the potential benefits of using automation in development and production environments by researchers and practitioners. Finally, answering \textbf{RQ4} is appropriate for researchers and practitioners to comprehend the potential open challenges for a seamless replication of development and production environments.

\section{Research Method}

\begin{figure}[!ht]
   \centering
     \includegraphics[scale = 0.6, clip = true, trim= 0pt 0pt 0pt 0pt]{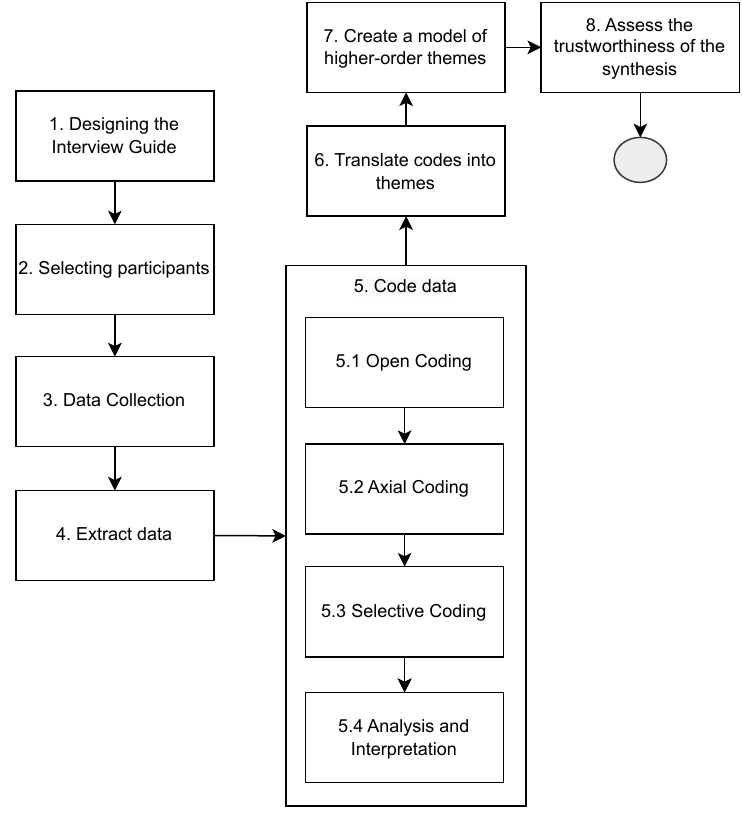}
   \caption{Research workflow.}
   \Description{Research workflow.}
   \label{fig:flowInterview}
\end{figure}

To answer our research questions, we conducted semi-structured interviews, followed by a Thematic Analysis procedure~\cite{cruzes2011ThematicAnalyse} to analyze our data. Figure~\ref{fig:flowInterview} presents an overview of our research method, which involves eight main steps: 1) Design of the interview protocol, 2) Participant selection, 3) Data collection, 3) extract data, 5) code data, 6) translate codes into themes, 7) create a model of higher-order themes, 8) assess the trustworthiness of the synthesis. In the following sections, we discuss the main steps in detail. 

\subsection{Designing the Interview Protocol}\label{sec:design}

The study uses semi-structured interviews with both closed-ended and open-ended questions to investigate how software companies handle configuration changes between development and production environments and deal with unforeseen problems~\cite{dearnley2005reflection}. With semi-structured interviews, we could also focus on different perspectives in different conversations according to the relevance of each theme for each context. Semi-structured interviews are also attractive to our context because we can even draw conclusions even with a few participants bringing detailed insight based on their experiences.

We design an interview guideline with key questions to the interviews, taking advantage of a generic interview protocol~\cite{leite2021organization} and other relevant work~\cite{Adams2015}. The interview questions address different themes, including: (1) development environment configuration, (2) production environment configuring, (3) incident handling, (4) strategies for mitigating the differences between environments, and (5) the effectiveness of the mitigation strategies.

The interview script comprises 4 parts. In the first part, we asked our interviewees \textbf{demographics} questions about their profile, career in the industry, role in the company, current project.
In the second part, we asked the interviewees about the \textbf{development environment}, how is it built?, is it local or cloud-based?, has it any automation?, are the development environment separated from each of other per project?.
In the third part, we asked the interviewees about the \textbf{production environment}, how is it built?, is it local or cloud-based?, has it any automation?, are the production environment separated from each of other per project?.
In the fourth part, we asked the interviewees about the \textbf{mitigation strategies} adopted in the company to mitigate the differences between the environments.

To evaluate the efficacy of our interview design, a pilot study was carried out with an emphasis on alignment with research objectives and possible areas for improvement. While the 61-minute pilot interview provided insightful information, no notable modifications were found. The primary round of interviews includes the pilot.

\subsection{Selecting participants}\label{sec:participants}

In order to understand the possible strategies to mitigate configuration differences between development and production environments, it is important to gather insights and perspectives from practitioners. As such, one prerequisite was that a participant candidate should be actively involved in DevOps-aligned workflows work and in an industrial context that has adopted IaC or should be implementing measures toward setting up IaC. A participant candidate should also have at least four years of working experience in IT, based on the Stack Overflow survey that informs that 48\% of respondents have less than ten years of experience in coding~\cite{StackOverflowSurvey23}. IaC is closely related to configuration management, as it involves defining and provisioning infrastructure resources programmatically and is widely adopted in DevOps and cloud-native environments. For this reason, we believe that participants who have expertise in IaC are more likely to have a solid understanding of configuration management principles and practices.

Interviewing experienced IT professionals, operations managers, and other stakeholders assists our study because they shed light on technical expertise, problem-solving skills, and industry knowledge while offering a variety of perspectives on current processes, practices, problems, and issues.

Eight of the ten close associates who were first invited to participate in the study responded, using a convenient sampling approach. Researchers used social media (Twitter, Telegram, and Discord) pertaining to DevOps, DevSecOps, SRE, and Ansible, as well as referrals from interviewees and coworkers, in order to expand the sample. While some invitees were interested but unable to make it, others did not respond. In the end, \totalParticipants professional participated in our study. All participants work for distinct companies with different workflows to deliver software. The demographic information regarding the participants is depicted in Table~\ref{table:demographicData}. We then opted to stop conducting interviews, given the number of participants we were able to recruit (more on this in Section~\ref{sec:limitations}). A saturation of the analysis was shown by the fact that no new categories were added, despite additional interviews confirming the ones that already existed.

\begin{table}[b]
\centering
\scriptsize
\begin{tabular}{@{}cccccl@{}}
\toprule
\textbf{\#} & \textbf{Gender} &  \textbf{Exp. in IT} & \textbf{Exp. in IaC} & \multicolumn{1}{c}{\textbf{Role}} & \textbf{Duration} \\ \midrule
P1  & Male   &  19y & 1y & Senior Software Developer        & 61m \\
P2  & Female &  11y & 2y & Senior Site Reliability Engineer & 33m \\
P3  & Male    & 19y & 4y &  Software Developer               & 29m \\
P4  & Male   & 5y  & 1y & DevOps Cloud Analyst             & 21m \\
P5  & Male    & 15y & 2y & Systems Analyst                  & 19m \\
P6  & Male    & 14y & 3y &  Software Developer               & 23m \\
P7  & Male    & 7y  & 2y &  Systems Analyst and Developer    & 27m \\
P8  & Male    & 17y & 2y & Senior Solutions Architect       & 37m \\
P9  & Male    & 7y  & 2y &  Software Developer               & 21m \\
P10 & Male    & 8y  & 1y &  Senior Tech Support Engineer     & 27m \\
P11 & Male    & 6y  & 4y &  IT Infrastructure Manager        & 28m \\
P12 & Male    & 14y & 4y & IT Manager                       & 41m \\
P13 & Female & 12y & 2y &  DevOps Cloud Analyst             & 29m \\
P14 & Male    & 14y & 1y &  Technical Lead                   & 22m \\
P15 & Male    & 16y & 3y &  IT Manager                       & 45m \\
P16 & Male    & 24y & 3y &  DevOps Tech Lead                 & 38m \\
P17 & Male    & 20y & 2y & Systems Analyst                  & 63m \\ \bottomrule
\end{tabular}
\caption{Demographic distribution of participants, including years of experience in IT and IaC. Participants' expertise varies, ensuring diverse perspectives on configuration management practices.}
\label{table:demographicData}
\end{table}

\subsection{Data collection}\label{sec:dataCollection}


In the interviews, the first author explained the purpose of the study, the role of the participant in the interview, and how we planned to use the data. Participants were explicitly instructed to avoid sharing personal or sensitive information. The interviews were conduced in Portuguese during the period of August 2021 and January 2023. In order to preserve anonymity and comply with ethical guidelines~\cite{Strandberg2019}, interviews were coded in Portuguese and only translated into English for manuscript preparation. The duration of the interviews ranged from 20 to 63 minutes, with an average of 33 minutes. When participants shared similar viewpoints or responded succinctly and consistently to simple questions, the sessions lasted less time. 

The study involved P1--P\totalParticipants participants with an average of 14 years in Information Technology and three years with IaC, gathered over nine hours and 32 minutes of overall interview data. After conducting the interviews, we transcribed each one of them for later analysis (Section~\ref{sec:analysis}).

\subsection{Analyzing the interviews}\label{sec:analysis}

We follow a Thematic Analysis procedure to analyze our data~\cite{cruzes2011ThematicAnalyse}. This procedure could be summarized as: Extract data, code data, Translate codes into themes, create a model of higher-order themes, assess the trustworthiness of the synthesis.

\subsubsection{Extract and Code data}\label{subsec:extractAndCodeData}

During this process, we created two artifacts for each interview: the \textbf{transcripts} and the \textbf{codes}. We also created one global artifact \textbf{the comparison sheet}. The first author analyzed the interview transcripts, while discussing pertinent codes with the last author. The majority of code where emerged inductively from the thematic analysis of our semi-structured interviews and influenced by prior work~\cite{myCodes}.

\begin{enumerate}
   \item \textit{Data Collection}. Semi-structured interviews were conducted with \totalParticipants experienced participants involved in the software industry to gather qualitative data.
   
   \item \textit{Data Transcription}. We listened to each audio record and manually transcribed it into text format, facilitating analysis and interpretation of the interview findings. We did transcribe the complete interview. However, we excluded meaningless noise~\cite{Adams2015}. For instance, we transcribed the following segment of a conversation with the participant P16: \textit{``We build the production environment the same way as the development environment. We run the same roles using Ansible but with different values for role variables in Ansible.''}
   
   \item \textit{Open Coding}. We conducted open coding by thoroughly reading the interview transcripts and identifying 32 initial codes that represented different strategies for managing configuration differences. For example:
       \begin{itemize}
            \item Code 1: Version control for configuration files
            \item Code 2: Environment Security Measures
            \item Code 3: Configuration management tools
            \item Code 4: Restricted Access Control
            \item Code 5: Infrastructure Automation Tools
            \item Code 6: Authorization Barriers
       \end{itemize}

    \item \textit{Axial Coding}. We performed axial coding to explore relationships and connections between the initial codes. This involved identifying sub-categories or relationships among the codes, resulting in 15 categories of codes. For example:
        \begin{itemize}
            \item Code 1: Automated Infrastructure Provisioning
                \begin{itemize}
                    \item Sub-category 1: Version control for configuration files
                    \item Sub-category 2: Configuration management tools
                    \item Sub-category 3: Infrastructure Automation Tools
                \end{itemize}
            \item Code 2: Access restriction between environments
                \begin{itemize}
                    \item Sub-category 1: Environment Security Measures
                    \item Sub-category 2: Restricted Access Control
                    \item Sub-category 3: Authorization Barriers
                \end{itemize}
        \end{itemize}

    \item \textit{Selective Coding}. We decided to keep 14 refined categories of codes because they are focusing on core categories that represent the main focus of the study. We filled the cells with concise statements, with columns representing participants and rows representing the summarized codes, as we show in Table~\ref{table:interviewCodes}. The code named ``Independence between projects'' was the one removed for not focusing on core categories.

    \item \textit{Analysis and Interpretation}. We analyzed and interpreted the coded data to draw conclusions and insights regarding the benefits and difficulties associated with different strategies for managing configuration differences between development and production environments in software development.
\end{enumerate}

\begin{table}[h!]
    \centering
    \small
    \begin{tabular}{@{}llc@{}}
    \toprule
    \multicolumn{1}{c}{\textbf{Codes}}                        & \textbf{\#}                     & \textbf{\%} \\ \midrule
    Consistency of data \\and settings between environments   & P1-P5,P8-P17                    & 88\%        \\
    Build Validation Mechanisms                               & P1-P3,P6-P13,P15-P17            & 82\%        \\
    Time-saving                                               & P1,P2,P4,P6-P8,P10-P14,P16,P17  & 76\%        \\
    Dependencies between \\environments along the pipeline    & P1,P5-P8,P10-P12,P14-P16        & 65\%        \\
    Well-defined scope of \\action between teams              & P1-P3,P9-P11,P13-P15            & 53\%        \\
    Infrastructure Manual \\Provisioning                      & P2,P3,P5-P7,P9,P11,P12,P15,P17  & 59\%        \\
    Access restriction \\between environments                 & P1,P3,P5,P7,P8,P12,P14-P16      & 53\%        \\
    Automated Infrastructure Provisioning                     & P1,P2,P4,P8,P10,P13,P14,P16,P17 & 53\%        \\
    Control of financial expenses                             & P1,P3,P4,P8,P14-P16             & 41\%        \\
    Technical debt                                            & P2,P6,P7,P9,P11,P15,P16         & 41\%        \\
    Lack of involvement \\with other teams                    & P1-P3,P7,P12,P13                & 35\%        \\
    Inconsistency of data \\and settings between environments & P1,P2,P11,P14                   & 24\%        \\
    Team communication mechanisms                             & P1,P2,P15,P17                   & 24\%        \\
    Monitoring Mechanisms                                     & P1,P2,P4                        & 18\%        \\ \bottomrule
    
    \end{tabular}
    \caption{Interview Codes with an average of occurrences.}
    \label{table:interviewCodes}
\end{table}

\subsubsection{Translate codes into themes}\label{subsec:translateCodesToThemes}

Thematic Analysis involves translating codes into themes, identifying patterns and connections among similar meanings. The author reviews codes and groups related codes into broader categories to encapsulate the data's content from Table~\ref{table:interviewCodes}. This comprehensive, iterative process ensures the themes accurately represent key elements. These themes were independently examined by the author and the last author to further increase credibility, providing a structured framework for organizing and interpreting data. The generated themes are listed below:

The theme \textbf{Access Control and Security} is centered on controlling access privileges, guaranteeing data security and protection, and making sure that only people with permission are allowed into environments. Which is associated with code: Access restriction between environments.

The theme \textbf{Data and Environment Management} focuses on preserving data consistency and integrity across different systems or process settings. Which is associated with codes: Consistency of data and settings between environments, Dependencies between environments along the pipeline, Inconsistency of data and settings between environments.

The theme \textbf{Infrastructure Provisioning} concentrates on configuring and managing software development and deployment infrastructure, which impacts quality assurance, consistency, dependability, data management, and validation in development, testing, and deployment environments. Which is associated with codes: Automated Infrastructure Provisioning, Infrastructure Manual Provisioning.        

The theme \textbf{Monitoring and Performance} focuses on monitoring systems that are essential for tracking system functionality, data management, validation, teamwork, infrastructure provisioning, access control, problem discovery, performance optimization, and system integrity. Which is associated with code: Monitoring Mechanisms.

The theme \textbf{Team Collaboration and Communication} highlights how crucial good teamwork and communication are to the successful provisioning of infrastructure, validation, and data and environment management. Which is associated with codes: Lack of involvement with other teams, Team communication mechanisms, Time-saving, Well-defined scope of action between teams.

The theme \textbf{Validation and Quality Assurance} ensures system quality, accuracy, and reliability, managing technological debt, and resolving data and environment consistency challenges. Which is associated with codes: Build Validation Mechanisms, Control of financial expenses, Technical debt.

\subsubsection{Create a model of higher-order themes}\label{subsec:higherOrderThemes}

Thematic Analysis involves synthesizing and organizing identified higher-order themes into a coherent structure, establishing relationships, identifying patterns, and identifying overarching concepts. This process helps identify broader, abstract categories and second-order themes, providing a conceptual framework for understanding key insights. It requires a reflective approach for a comprehensive understanding. We grouped the previous themes into two new ones: Data Governance and Management and Operational Efficiency and Effectiveness.

The second-order theme, \textbf{Data Governance and Management}, focuses on how businesses can support efficient operations and performance monitoring by implementing effective data governance and management procedures that ensure data availability, correctness, integrity, and security. Which is associated with themes: Access Control and Security, Data and Environment Management, Validation and Quality Assurance.

\begin{figure}[!ht]
   \centering
     \includegraphics[scale = 0.5, clip = true, trim= 0pt 0pt 0pt 0pt]{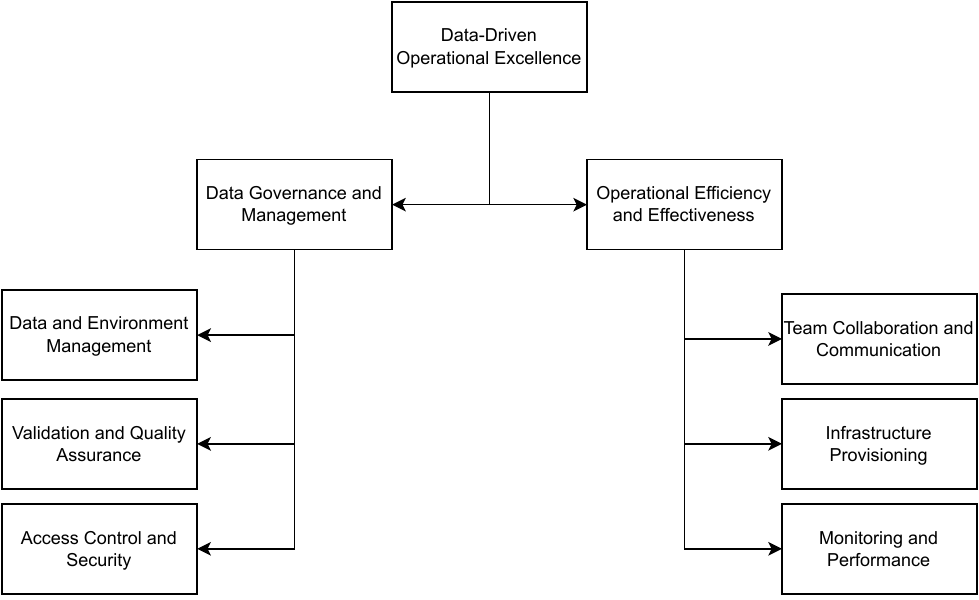}
   \caption{Thematic model of strategies for managing configuration differences.}
   \Description{Model of higher-order themes.}
   \label{fig:higherOrderThemes}
\end{figure}

The second-order theme, \textbf{Operational Efficiency and Effectiveness} emphasizes effective data governance, efficient decision-making, and access to high-quality data for better operational success. Which is associated with themes: Infrastructure Provisioning, Monitoring and Performance, Team Collaboration and Communication.
By combining Operational Efficiency, Effectiveness, Data Governance, and Management into a single topic known as Data-Driven Operational Excellence, we emphasize the mutually beneficial and related nature of these concepts. The final structure can be seen in the Figure~\ref{fig:higherOrderThemes}.

The theme \textbf{Data-Driven Operational Excellence} emphasizes how crucial it is to use efficient data governance and management procedures that improve an organization's operational effectiveness. Which is associated with themes: Data Governance and Management, Operational Efficiency and Effectiveness.

\subsubsection{Assess the trustworthiness of the synthesis}\label{subsec:AssessTrustworthinessSynthesis}

As every research project should be assessed in terms of the methodologies utilized, and our research findings should be as reliable or feasible. The quality and quantity of the evidence that a synthesis is based on determines how reliable the results are. The credibility~\cite{cruzes2011ThematicAnalyse} of our qualitative research could be ensured by several crucial procedures in the context of the themes identified. To build trustworthiness thoroughly data analysis is first and foremost necessary. To correctly represent the viewpoints and experiences of the participants, we thoroughly and methodically evaluate the data in search of patterns, correlations, and themes. The data was independently examined by several researchers to further increase credibility. The interpretations of each researcher might then be contrasted and discussed, assisting in the detection of potential biases and ensuring that the themes are firmly rooted in the data rather than being impacted by personal viewpoints. Furthermore, requesting peer review was a crucial step in building credibility. The accuracy of the themes and their alignment with the participants' viewpoints were validated and confirmed through the peer debriefing procedure.

\section{Research Results}

This section details our research results, organized in terms of the research questions.

\subsection{RQ1 -- What actions are taken?}

Because the original research themes were inadequate, a fresh round of derived themes was required in order to explore second-order themes and offer more thorough insights into the idea of \textbf{strategy}. It attempts to reduce the disparities between development and production environments by putting procedures, tools, and configurations into place. The last author examined the new \textbf{strategy themes} that the first author separately developed for comparison and debate without making any more changes or revisions. The Table~\ref{table:typesStrategies} summarizes the strategies that emerged during our research.

The \textbf{Automated Deployment Pipeline} strategy is often combined with the CI/CD system, which consists of automating the construction, testing, and deployment processes. Teams may more easily and swiftly distribute updates and work together more efficiently because of its automation of the deployment process, reduction of human error, and assurance of appropriate testing. For example, participant P4 declared: \textit{``In my experience,  I just went to GitHub Actions and put a dropdown menu where we can choose which environment it will generate the build for, and it already publishes this kind of change.''}. Nevertheless, this strategy has a few disadvantages. Because of the frequent release of new software versions and the discovery of bugs, pipelines are complex, difficult to manage, and require regular maintenance. By using CI/CD tools like Jenkins, GitLab CI/CD, or GitHub Actions to create an Automated Deployment Pipeline, teams may automate build and deployment procedures. Use parameterized builds or environment variables to specify configurations unique to a given environment. Ensure rollback methods by using feature flags or blue-green deployments. Use tools like Terraform or Ansible to automate infrastructure provisioning so that environments are consistent.

The \textbf{Build Profiles Management} strategy uses build tools to define and maintain software settings for various environments, guaranteeing efficiency and consistency. By automating the procedure, it lowers errors, inconsistencies, and human error while increasing the speed and efficiency of configuration modifications. For example, Apache Maven structure this strategy into profiles that contain all the application configuration per environment. Once specified the profile, it selects and merges the configurations at build time, guaranteeing that all configurations will be there. For example, participant P15 stated: \textit{``We have solutions based in build files that we organize using named build profiles referring to each environment, and according to the type of build, one of these profiles will be used.''}. Nevertheless, this strategy has a few disadvantages. It requires specialized knowledge and technical expertise. It can need regular maintenance and support, which might put a heavy burden on IT workers. It imposes a complete rebuild of the application to apply changes in configurations. To ensure efficient build profile management, teams may use profile-based configuration tools like Apache Maven or Gradle to define settings per environment. Maintain separate environment profiles (e.g., dev.properties, staging.properties, prod.properties) with a mechanism to load the appropriate configuration at runtime. Store configurations in a centralized system such as Kubernetes ConfigMaps, AWS Parameter Store, or HashiCorp Vault.

The \textbf{Configuration Management Plan (CMP)} strategy helps the SCM be managed holistically during the development and deployment phases. It entails recording, monitoring, and guaranteeing conformity to setup guidelines. It enhances problem-solving and teamwork by guaranteeing consistency, monitoring changes, and adhering to regulatory compliance standards. For instance, these plans in the Ansible tool are called \textit{Playbooks} and the elements of this plan are called \textit{tasks}. As a result, teams could quickly assess the present state of the environment and decide what adjustments will be necessary. For example, participant P2 announced: \textit{``We use the same Terraform for the different environments, just changing the parameters according to the environment because the environment must always have the same structure.''}. Also, participant P1 declared: \textit{``We have a development environment that is created automatically through Puppet. The team configures the creation and updates Puppet scripts through the parameters for the new environment.''}. Yet, this strategy has several disadvantages. Complex plans are difficult to implement successfully and adjust to changing needs because they require specialized resources, time-consuming implementation, rigidity, and complexity. To develop a CMP that works well for setting configuration baselines and using version control (like Git) to enforce them. Use IaC technologies such as Chef, Ansible, or Puppet to automate configuration enforcement. Use compliance-as-code solutions such as AWS Config or Open Policy Agent (OPA) to enforce ongoing configuration validation. With the use of logging tools such as Splunk, Prometheus, or ELK Stack, audit and track configuration changes.

The \textbf{Database Table Configuration} strategy managed and configured database tables for a software application to store data consistently and effectively. It is centralized for communication and cooperation, audited for security and regulatory compliance, and appropriate for large companies with complex infrastructure. For example, participant P15 declared: \textit{``All of our services read all settings via the database.''}. However, this strategy has several disadvantages. It may result in inconsistencies and possible errors by increasing data complexity, increasing performance overhead, and limiting integration with other systems. To maintain consistent database configurations across environments to use database schema migration tools such as Flyway or Liquibase to track changes systematically. Enforce environment-specific database configurations using separate .env files or secrets management tools. Automate backups and restores to prevent inconsistencies between development and production databases. Use database seeding strategies to create realistic test data in staging environments without exposing sensitive production data.

Curiously, 35\% of the participants of our study still use a \textbf{Manual Configuration} strategy involves managing configurations manually. Manually maintaining setups offers flexibility, hands-on experience, and knowledge, but can be time-consuming, inconsistent, inefficient, and not easily scalable. In large businesses, it can take a lot of time and might be inconsistent because of oversight or human mistake. Although automated solutions can be more effective, manual administration can be difficult in companies that are expanding quickly. For example, participant P5 declared: \textit{``Due to our lack of knowledge, we manually change the environments that the build process generates.''}. To minimize risks while using manual configurations to maintain a configuration checklist to ensure consistency. Document changes thoroughly to prevent errors and facilitate troubleshooting. Use scripts where possible (e.g., shell scripts) to automate repetitive tasks.

The \textbf{Operating System Configuration} strategy includes modifying system settings, controlling user access, establishing security protocols, and allocating resources as efficiently as possible. It keeps constant configuration data, is portable across systems, and may be programmed or scripted to automate large-scale modifications. For example, participant P12 affirmed: \textit{``We use a checklist to manually confirm the changes that need to be made in each environment. Usually, they are variables that we put in the operating systems the application will look for to know its current configuration.''}. Nonetheless, this strategy has disadvantages. It is a possible security risk because anyone with sufficient privileges can access it in plain text. Also, managing these variables can become complex, especially in large organizations with multiple systems and environments without an automation tool. To ensure consistent OS-level configuration, teams could automate OS configuration using tools like Ansible, Puppet, or SaltStack. Use environment variables to standardize settings across different environments. Apply configuration validation tools to detect misconfigurations before deployment.

The \textbf{Platform as a Service (PaaS) deployment} strategy simplifies infrastructure administration by managing software applications through a local or cloud-based platform. It provides scalable, pre-configured infrastructures that enable businesses to grow as needed and simplify settings for non-technical employees. Additionally, PaaS has integrated automation capabilities that lower the possibility of human error and increase cost-effectiveness. For example, participant P16 declared: \textit{``We always select the same code for different environments, only changing the Dokku variables.''}. As another example, participant P14 declared: \textit{``We work with configuration files with environment variables. These variables are provided via Kubernetes Secrets and Configmaps for each application.''}. Regardless, this strategy has several disadvantages. The absence of explicit standards may cause organizations to struggle with configuration security and customisation. Reliability determines the provider's quality of service, and moving providers can be challenging, possibly securing an organization's place with a specific supplier.  To effectively deploy applications using PaaS to choose a provider (e.g., AWS Elastic Beanstalk, Google App Engine, Azure App Service, Heroku). Define environment variables and configuration settings within the PaaS platform to ensure consistency. Utilize containerized deployment models such as Docker and Kubernetes for better portability. Implement auto-scaling rules to dynamically adjust resources based on traffic patterns.

the \textbf{Web Server Configuration} strategy to manage security measures, server settings, and access credentials. It has benefits including rapid updates, scalability, and ease of accessibility. For example, Apache Tomcat has a feature named JNDI Datasource that helps with this kind of configuration. For example, participant P5 affirmed: \textit{``Today, with the use of Spring Boot, we had more facilities because now we can place an application and configure this application through either an external file or a datasource, which resides in each environment.''}. Nonetheless, this strategy has several disadvantages. For larger companies without automation capabilities, storing sensitive configuration data on a remote server can be time-consuming and present security risks. It can be difficult to maintain the server and its settings. To properly configure web servers for optimal performance and security to use configuration management tools (e.g., Apache Tomcat, Nginx, HAProxy) to ensure consistent settings. Automate deployment of server configurations using scripts or orchestration tools. Use HTTPS and authentication mechanisms to secure server configurations against unauthorized access.

\begin{table}[!ht]
\centering
\footnotesize
\begin{tabular}{@{}llcc@{}}
\toprule
\multicolumn{1}{c}{Types of Strategy} & \# & \%   & Ref  \\ \midrule
Automated Deployment Pipeline         & P1,P2,P4,P16       & 24\% & ~\cite{poth2018deliver} \\
Build Profiles Management             & P6,P15             & 12\% & ~\cite{rahman2017factors} \\
Configuration Management Plans        & P1,P2              & 12\% & ~\cite{krishna2018managing} \\
Database Table Configuration          & P9,P15             & 12\% & ~\cite{Nguyen2006} \\
Manual Configuration                  & P5,P6,P10-P12,P17  & 35\% & ~\cite{perera2018continuous}\\
Operating System Configuration        & P3,P12             & 12\% & ~\cite{zeng2014managing}\\
PaaS Deployment                       & P8,P13,P14,P16     & 24\% & ~\cite{keller2010platform}  \\
Web Server Configuration              & P5,P7,P9,P17       & 24\% & ~\cite{ApacheJNDIDatasource}\\ \bottomrule
\end{tabular}
\caption{Summary of strategy adoption across participants, categorized by configuration management approaches.}
\label{table:typesStrategies}
\end{table}

\subsection{RQ2 -- Are these team actions sufficient?}

The themes created for RQ1 first appeared valuable, but they were unable to capture important elements of RQ2, which called for evaluating the efficacy of mitigating techniques. The researchers made the decision to develop new codes specifically for RQ2 because the preexisting themes were not intended to convey sufficiency. In order to categorize strategies as \textbf{Sufficient}, \textbf{Insufficient}, \textbf{Partially Sufficient}, and \textbf{Not Specified}. These measurements served as qualitative indicators to characterize participants' views on the adequacy and effectiveness of the strategies, delivering a structured and concise way to present the insights gleaned from the analysis.

The ``Sufficient'' category participants think that their existing infrastructure management procedures adequately handle configuration differences between development and production environments, guaranteeing dependable and fluid system operation. While acknowledging that certain aspects require work, participants in the ``Partially Sufficient'' group think that current infrastructure management approaches adequately handle configuration changes. Despite their company's procedures, tools, and technologies, participants in the ``Insufficient'' category feel that their existing infrastructure management techniques are not enough to handle configuration changes. The participants who are unfamiliar with or lack practical knowledge with current infrastructure management approaches fall into the ``Not Specified'' group. Table~\ref{table:perceptionMitigationStrategies} summarizes the result of the interviews conducted with \totalParticipants participants. 

Despite the efficiency of the current strategy, participants reported disappointment with the strategies and identified areas for development when combined with other automation technologies. For example, participant P7 agrees with the adopted strategy, but there are still some concerns informed: \textit{``These meet our needs 100\%, but we are dependent on previously running the database routines applied manually in the new environments, as our JPA (Java Persistence API) configurations do not automatically create tables and other database artifacts when we place the new build at production.''}. Furthermore, participant P10 declared: \textit{``The current strategy is not enough because there are better ways of configuring this, one of them is via Ansible, and I am already working on it to make the configuration even more automatic.''}. Also, participant P8 stated: \textit{``No, because sometimes we have to manually interfere for that specific situation to be resolved more quickly, once resolved this change will be incorporated into our scripts.''}.

\begin{table}[!ht]
\centering
\footnotesize
\begin{tabular}{@{}llc@{}}
\toprule
\multicolumn{1}{c}{Perception} & \# & \% \\ \midrule
Sufficient                     & P1,P7,P13,P14,P17      & 29\% \\
Partially Sufficient           & P2,P3,P9,P15,P16       & 29\% \\
Insufficient                   & P4,P8,P10,P11          & 24\% \\
Not Specified                  & P5,P6,P12              & 18\% \\ \bottomrule
\end{tabular}
\caption{Average of the Perception of the satisfaction by employed mitigation strategies in the companies.}
\label{table:perceptionMitigationStrategies}
\end{table}

\subsection{RQ3 -- What are the perceived benefits?}

We consider the interview second-order themes from Figure~\ref{fig:higherOrderThemes} and the interview codes from Table~\ref{table:interviewCodes}, because when using themes instead of codes in thematic analysis provides a higher level of abstraction and organization to the analysis process. While codes represent specific segments of data, themes offer a broader and more conceptual grouping of related codes. Each of them represents a specific aspect or advantage related to the implementation of certain practices or mechanisms.

The theme \textbf{Data and Environment Management} deals with the need to ensure consistency of settings across contexts and to protect data integrity. The related code found is \textbf{Consistency of data and settings between environments} ensures that the data and configuration settings are kept consistent across different environments to avoid discrepancies and ensure smooth operation. For example, participant P8 mentioned: \textit{``The great benefit is when someone who is going to join our development team, everything they need to run our application is automated via docker-compose. Consequently, it makes working on the project a lot easier. Another benefit is that maintaining the settings of our application has become easier for us.''}.

The theme \textbf{Infrastructure Provisioning} focuses on the methods and resources needed to configure and oversee the infrastructure needed for software development and implementation, a process known as infrastructure provisioning. The related code is \textbf{Automated Infrastructure Provisioning} utilizes automation tools and processes to provision and configure the required infrastructure resources for software development and deployment, streamlining the setup and management. For example, participant P3 declared: \textit{``Therefore, today I want to change a server, I can go to the cloud platform and select it, and I will pay more, but I will have a server in a few minutes.''}.

The theme \textbf{Monitoring and Performance} highlights the need of monitoring systems for tracking and assessing the functionality and condition of the application or system. The related code is \textbf{Monitoring Mechanisms} implementing systems and tools to monitor the performance, availability, and health of software applications and infrastructure components. For example, participant P2 mentioned: \textit{``We have alerts when it makes sense. We have a member rotation in our team. So that there is always someone keeping an eye on it. If there is a problem, we already look at it, or if it is a scenario that even testing has escaped us.''}.

The theme \textbf{Team Collaboration and Communication} emphasis on effective teamwork and communication. The tasks assigned to each team must be clearly defined, cooperation must be seamless, and teams cannot be isolated from one another. The related code is \textbf{Time-saving}. This code refers to the perceived reduction in time and effort required to complete a task or achieve a goal. For example, participant P2 shared: \textit{``Of course, our job as DevOps is to automate as much as we can because when we automate, more time is left for us to improve our infrastructure, automate other things, and improve processes, sometimes there is an old machine that we need to update or there is a bad process, and we need to improve it or there is a part of the infrastructure that does not yet have Terraform, and we need to apply it. The idea is that we have more time to do more things and improve the infrastructure in a general context.''}.

Finally, our study also reveals the theme \textbf{Validation and Quality Assurance} includes validation processes and technical debt management. The related code is \textbf{Build Validation Mechanisms} implementing mechanisms to validate the integrity and correctness of software builds, ensuring that they meet quality standards and are suitable for deployment. For example, participant P8 mentioned: \textit{``We use two container testing approaches. The first is the smoke test to test some application functionality. The second is the integration test that evaluates the communication between microservices.''}.

\subsection{RQ4 -- What are the perceived difficulties?}

Similarly to RQ3, we consider the interview second-order themes from Figure~\ref{fig:higherOrderThemes} and the interview codes from Table~\ref{table:interviewCodes} to answer this question.

The theme \textbf{Access Control and Security} manage and regulate access privileges to different settings. It is related to the code \textbf{Access restriction between environments} implementing measures to restrict access and control permissions between different environments (e.g., development, production) to maintain security and prevent unauthorized actions. For example, participant P12 mentioned: \textit{``That depends on the client's business model. The team can host the application or delivery it to the client's team to host in their production environment due to access restrictions.''}.

The theme \textbf{Data and Environment Management} deals with the need to ensure consistency of settings across contexts and to protect data integrity. The first related code is \textbf{Dependencies between environments along the pipeline} that identifies and manages dependencies between various environments within the software development and deployment pipeline to ensure smooth progression and avoid bottlenecks. For example, participant P1 declared: \textit{``Yes, we have a specific environment on the developer's machine because it is where we build our applications. However, we, unfortunately, depend on another environment configured with permissions to work correctly. We call it the general development environment. We will have an approved certificate, which gives us authenticity with a secret key. We will have other subsystems that connect and support our applications.''}. The second related code is \textbf{Inconsistency of data and settings between environments} that highlights the presence of discrepancies and inconsistencies in data and configuration settings across different environments, which can lead to issues and errors. For example, participant P11 shared: \textit{``We have difficulty automating in a standardized way due to the diversity of operating systems. We can not guarantee that all have the same configuration or even data.''}. Another participant mentioned: \textit{``In my previous company, I always had to go to production to make a copy of a piece of data in the database, and I had to copy the technical material of the production environment to load it into the development environment or vice versa. It does not always work for several reasons. For example, a dependency on the development environment that did not have in production or vice versa.''}.

The theme \textbf{Infrastructure Provisioning} focuses on the methods needed to configure and oversee the infrastructure needed for software development and implementation, a process known as infrastructure provisioning. The related code is \textbf{Infrastructure Manual Provisioning} does manual provision or setting up of the necessary infrastructure resources (e.g., servers, databases) for software development and deployment. For example, participant P9 declared: \textit{``Since the process is done manually and without standardization, we have issues with our servers, such as memory issues or crashes. Both are mysterious and only resolve themselves after a restart.''}

The theme \textbf{Team Collaboration and Communication} emphasis on effective teamwork and communication. The tasks assigned to each team must be clearly defined, cooperation must be seamless, and teams cannot be isolated from one another. The first related code is \textbf{Lack of involvement with other teams} denotes a lack of involvement or interaction with other teams, potentially hindering collaboration and coordination efforts. For example, participant P3 mentioned: \textit{``So it depends a lot on the solution of the product, what is its purpose, right? So, this has a series of rules here that I do not know how to detail what the rules are to decide where it goes.''}. The second related code is \textbf{Team communication mechanisms} the processes and tools in place to facilitate effective communication and collaboration among team members involved in software development and deployment. For example, participant P15 mentioned: \textit{``The development team should prepare a deployment plan, and from there, the developer should open a ticket to the production team, and they are the ones who should put the build into production. But sometimes we have to improvise due to problems in our communication''}. The last related code is \textbf{Well-defined scope of action between teams} clearly defining and communicating the areas of responsibility and authority for different teams involved in the software development and deployment process to avoid conflicts or duplication of efforts. For example, participant P13 mentioned: \textit{``Each team chooses who will be responsible for deploying the solution in production, but sometimes we have to be waiting for a DevOps.''}.

Finally, the theme \textbf{Validation and Quality Assurance} includes validation processes and technical debt management. The first related code is \textbf{Control of financial expenses} implementing strategies and mechanisms to monitor and control the financial costs associated with software development and deployment activities. For example, participant P4 mentioned: \textit{``The machines in AWS for development are less elaborated than those for production due to cost-savings. We also turn off these development machines during the night because they will not be in use. This impact to manage at least two types of machines.''}. The second related code is \textbf{Technical debt} refers to the accumulated software development work that needs to be addressed or improved in the future due to shortcuts, temporary solutions, or suboptimal code. For example, participant P10 declared: \textit{``Therefore it is a matter of lack of time the priorities always end up appearing, and since some of them are not that hard like creating a database user, we are used to it manually. We prefer to throw it under the carpet, but we know we need to do this, and we already have ideas for this. For example, building a CLI or a bot they call ChatOps or building our own Slack Bot that can help, we have the ideas but could not implement them due to a series of other issues we have as a priority.''}.

\section{Discussion}

According to the interviews findings, one of the biggest problems in software development is the difference between the development and production environments. IT specialists and operations managers were among the participants who emphasized the difficulties in handling configuration variations. ``Manual Configuration'' (35\%) is still widely used, suggesting that consistency issues persist. According to ``Automated Deployment Pipelines'' (24\%), automation's benefits are becoming more widely acknowledged. Other strategies that represent a variety of methods include ``Web Server Configuration'', ``Automated Deployment Pipelines'', and ``PaaS Deployment'' (each 24\%). Proactive planning is evident in ``Build Profiles Management' and ``Configuration Management Plans'' (12\%). Unfortunately, there is no one-size-fits-all answer, as evidenced by the range of strategies, and organizations are investigating all of their options to determine which one best meets their unique requirements.
Another main finding is that some participants are not happy with the way they currently handle configuration changes between development and production systems. This dissatisfaction implies that current tactics are inadequate and produce less-than-ideal results. ``Automated Deployment Pipelines'' (24\%) provide advantages but might not completely address configuration issues, whereas ``Manual Configuration'' (35\%) is popular but can be laborious and prone to mistakes. Configuration technical debt is caused by strategy dissatisfaction which is brought on by ineffective practices and shortcuts that raise maintenance and development expenses. Inconsistencies, inefficiencies, and operational difficulties can result from poorly maintained configurations. In order to pay off this debt, more research and better business plans are needed. Taking proactive measures include investing in configuration management systems that are suited to the particular needs and complexity of each software project, automating processes, and streamlining processes.
The strategies companies adopt to deal with the differences between development and production environments are revealed via thematic analysis. Researchers and practitioners can enhance practices and decision-making by comprehending these patterns, which may enhance software development and deployment project outcomes.

\subsection{Implications}
The implications of this paper can be different for industry and academia. For industry, this study provides a catalog of strategies about how to handle configuration differences across software development and production environments, facilitating well-informed decision-making and enhancing system reliability. It also acts as a standard by which to compare industries. For the researchers, by looking at organizational strategies for handling configuration discrepancies, this work advances software development and management research while also directing future studies and serving as a resource for practitioners and students.

\subsection{Limitations}\label{sec:limitations}

There are restrictions and validity concerns associated with the study’s limited sample size. The scope of the investigation may have prevented the transcriptions of the tenth participant from producing new codes. Seven more subjects were interviewed after the tenth to establish that there were no new codes as well as confirm validity.
Participants in the study were varied in terms of their backgrounds, levels of experience, and viewpoints. However, the paper did not initially provide a comprehensive explanation of how factors such as domain, team size, and DevOps maturity might influence the adoption or suitability of each strategy. For example, we observed that smaller teams often rely on lightweight or manual approaches due to resource limitations, while larger teams tend to adopt structured configuration management plans and automation pipelines. These contextual differences may lead to different research findings in other settings.
To support replication and transparency, we encourage future work to report findings stratified by team size, domain, and organizational practices, making it easier to generalize or compare results across studies. This clarification improves the scientific rigor of our contribution and enables better alignment with practical realities across diverse organizations.
Additionally, the results could be biased by the participants’ seniority. The study was also unable to bring in more women, suggesting that more gender-diverse participants are needed for future studies. The validity of the study relied on how well the researchers recorded and interpreted the codes; any errors or misinterpretations could have an impact on the findings, which were avoided by meticulously looking over the transcriptions and coding processes.
Although this study provides insightful information on the viewpoints of experts, its usefulness is limited by its sample size, scope, and potential biases. Findings may be supported or contradicted by more data.

\section{Related Works}

Numerous earlier studies that empirically looked into the adoption, flaws, or difficulties of IaC connect to our work. 

Guerriero~\cite{Guerriero2019}, adopting the data coming from 44 semi-structured interviews to senior developers of as many companies, they shed light on the state of the practice in the adoption of IaC and the key software engineering challenges in the field. Particularly, how practitioners adopt and develop IaC.
Rahman~\cite{Rahman2018}, conducted study on the challenges of developing IaC, especially in configuration management tools. To help IaC engineers, they explored Stack Overflow for questions that programmers regularly asked. Again, queries related to puppets were prioritized in this context. They used qualitative analysis to identify the three most common question categories as being (i) syntax errors, (ii) provisioning instances, and (iii) assessing Puppet’s feasibility to accomplish certain tasks.
Rahman~\cite{Rahman2019} also provided a list of seven security smells in IaC. These were obtained from a qualitative analysis of Puppet scripts found in open-source repositories. The specified smells comprise: (1) granting admin privileges by default, (2) empty passwords, (3) hard-coded secrets, (4) invalid IP address binding, (5) suspicious comments (such as 'TODO' or 'FIXME'), (6) use of HTTP without TLS, and (7) use of weak cryptography algorithms. However, this is again restricted to Puppet scripts and not all smells are generalizable to other languages or tools.
Nazário~\cite{myGreyWork}, conducted a Rapid Review of Grey Literature to extract nine strategies for mitigating configuration differences, including security aspects like access credentials. They synthesize industry knowledge from secondary sources. While our work is empirically grounded, the methodological difference highlights practitioner insights vs. literature-based best practices, making them complementary approaches. Integrating both could provide a more comprehensive view of configuration management strategies.

Finally, Hezaveh~\cite{RahmanHezaveh2019} identify four main topics in the research areas surrounding the field of IaC through a Systematic Literature Review (SLR). These topics are (i) framework/tool for IaC; (ii) use of IaC; (iii) empirical studies related to IaC; and (iv) testing in IaC. They inferred that while several studies exist on framework and tools, research in the context of IaC defects and security flaws is still at its early stages.
The outcomes of the SLR are in line with the research presented in this paper: indeed, the state of research and practice in IaC is still immature, which calls for additional empirical and industry-focused studies, such as the one carried out throughout this work. Besides, not all the related works cover cataloging the strategies employed for managing configuration differences between development and production environments.

\section{Conclusion}

In this work, we conducted \totalParticipants semi-structured interviews with IT professional. In these interviews, we asked our interviewees questions about the development and the production environment, besides questions about the mitigation strategies adopted in the company to mitigate the differences between these environments. By analyzing the interviews, we developed a comprehensive catalog of \totalStrategies strategies used by companies to manage configuration differences and ensure that their systems are running smoothly and reliably. We observed three main findings.

\begin{itemize}
    \item The first, a detailed catalog of \totalStrategies strategies adopted by software companies to effectively mitigate the configuration differences between the development and production environments. 
    \item The Second, a more in-depth understanding of how these strategies contribute to reducing the risk of configuration-related issues and guaranteeing the smooth and reliable operation of software systems. 
    \item The third, their gain insight into the practices, processes, challenges, and issues faced by IT experts and other stakeholders involved in the management and deployment of systems.
\end{itemize}

However, these strategies should be customized to meet the unique demands of each hard, and there is no one-size-fits-all approach to reducing configuration variances. Effectiveness and adaptability depend on regular evaluation and updates. Software development and IT operations can be greatly impacted by an understanding of these strategies. For example, a large company with strict compliance demands may emphasize Configuration Management Plans and Access Control. At the same time, a startup may focus on PaaS Deployment and Automated Pipelines to be agile. A proper understanding of when and how to tailor each strategy can significantly influence both software development as well as IT operations. Rather than merely restating known practices, our study organizes them into a thematic model of Data-Driven Operational Excellence, capturing the interplay between technical, organizational, and economic factors. These findings offer fresh insights for both researchers and practitioners seeking to understand and improve the operational reality of configuration management under DevOps.

\begin{acks}
We thank the reviewers for their helpful comments. This work is partially supported by CNPq (404940/2024-2, 420406/2023-9, 442779/2023-2, 444802/2024-0).
\end{acks}

\bibliographystyle{ACM-Reference-Format}
\bibliography{sample-base}


\begin{thebibliography}{29}


\ifx \showCODEN    \undefined \def \showCODEN     #1{\unskip}     \fi
\ifx \showISBNx    \undefined \def \showISBNx     #1{\unskip}     \fi
\ifx \showISBNxiii \undefined \def \showISBNxiii  #1{\unskip}     \fi
\ifx \showISSN     \undefined \def \showISSN      #1{\unskip}     \fi
\ifx \showLCCN     \undefined \def \showLCCN      #1{\unskip}     \fi
\ifx \shownote     \undefined \def \shownote      #1{#1}          \fi
\ifx \showarticletitle \undefined \def \showarticletitle #1{#1}   \fi
\ifx \showURL      \undefined \def \showURL       {\relax}        \fi
\providecommand\bibfield[2]{#2}
\providecommand\bibinfo[2]{#2}
\providecommand\natexlab[1]{#1}
\providecommand\showeprint[2][]{arXiv:#2}

\bibitem[Adams(2015)]%
        {Adams2015}
\bibfield{author}{\bibinfo{person}{William~C. Adams}.} \bibinfo{year}{2015}\natexlab{}.
\newblock \bibinfo{booktitle}{\emph{Conducting Semi-Structured Interviews}}.
\newblock \bibinfo{publisher}{John Wiley \& Sons, Ltd}, \bibinfo{address}{111 River Street Hoboken, NJ}, Chapter~19, \bibinfo{pages}{492--505}.
\newblock
\showISBNx{9781119171386}
\href{https://doi.org/10.1002/9781119171386.ch19}{doi:\nolinkurl{10.1002/9781119171386.ch19}}
\showeprint{https://onlinelibrary.wiley.com/doi/pdf/10.1002/9781119171386.ch19}


\bibitem[Bartusevics(2013)]%
        {bartusevics2013methodology}
\bibfield{author}{\bibinfo{person}{Arturs Bartusevics}.} \bibinfo{year}{2013}\natexlab{}.
\newblock \showarticletitle{A Methodology for Model-Driven Software Configuration Management Implementation and Support}. In \bibinfo{booktitle}{\emph{proceedings of the international scientific conference}}. [LLU], \bibinfo{publisher}{LLU}, \bibinfo{address}{627 Davis Drive Suite 300 Morrisville, NC 27560}, \bibinfo{pages}{260--266}.
\newblock


\bibitem[Couch et~al\mbox{.}(2010)]%
        {couch2010troubleshooting}
\bibfield{author}{\bibinfo{person}{Alva~L Couch}, \bibinfo{person}{Mark Burgess}, {and} \bibinfo{person}{AS Cfengine}.} \bibinfo{year}{2010}\natexlab{}.
\newblock \showarticletitle{Troubleshooting with human-readable automated reasoning}. In \bibinfo{booktitle}{\emph{Proceedings of LISA’10: 24th Large Installation System Administration Conference}}. \bibinfo{publisher}{none}, \bibinfo{address}{none}, \bibinfo{pages}{239}.
\newblock


\bibitem[Cruzes and Dyba(2011)]%
        {cruzes2011ThematicAnalyse}
\bibfield{author}{\bibinfo{person}{Daniela~S Cruzes} {and} \bibinfo{person}{Tore Dyba}.} \bibinfo{year}{2011}\natexlab{}.
\newblock \showarticletitle{Recommended steps for thematic synthesis in software engineering}. In \bibinfo{booktitle}{\emph{2011 international symposium on empirical software engineering and measurement}}. IEEE, \bibinfo{publisher}{none}, \bibinfo{address}{none}, \bibinfo{pages}{275--284}.
\newblock


\bibitem[Dearnley(2005)]%
        {dearnley2005reflection}
\bibfield{author}{\bibinfo{person}{Christine Dearnley}.} \bibinfo{year}{2005}\natexlab{}.
\newblock \showarticletitle{A reflection on the use of semi-structured interviews}.
\newblock \bibinfo{journal}{\emph{Nurse researcher}} \bibinfo{volume}{13}, \bibinfo{number}{1} (\bibinfo{year}{2005}), \bibinfo{pages}{1}.
\newblock


\bibitem[Ebert et~al\mbox{.}(2016)]%
        {ebert2016devops}
\bibfield{author}{\bibinfo{person}{Christof Ebert}, \bibinfo{person}{Gorka Gallardo}, \bibinfo{person}{Josune Hernantes}, {and} \bibinfo{person}{Nicolas Serrano}.} \bibinfo{year}{2016}\natexlab{}.
\newblock \showarticletitle{DevOps}.
\newblock \bibinfo{journal}{\emph{Ieee Software}} \bibinfo{volume}{33}, \bibinfo{number}{3} (\bibinfo{year}{2016}), \bibinfo{pages}{94--100}.
\newblock


\bibitem[Fitzgerald and Stol(2017)]%
        {fitzgerald2017continuous}
\bibfield{author}{\bibinfo{person}{Brian Fitzgerald} {and} \bibinfo{person}{Klaas-Jan Stol}.} \bibinfo{year}{2017}\natexlab{}.
\newblock \showarticletitle{Continuous software engineering: A roadmap and agenda}.
\newblock \bibinfo{journal}{\emph{Journal of Systems and Software}}  \bibinfo{volume}{123} (\bibinfo{year}{2017}), \bibinfo{pages}{176--189}.
\newblock


\bibitem[Foundation(2018)]%
        {ApacheJNDIDatasource}
\bibfield{author}{\bibinfo{person}{The Apache~Software Foundation}.} \bibinfo{year}{2018}\natexlab{}.
\newblock \bibinfo{title}{Apache Tomcat 8 (8.0.53) - JNDI Datasource HOW-TO}.
\newblock \bibinfo{howpublished}{\url{https://tomcat.apache.org/tomcat-8.0-doc/jndi-datasource-examples-howto.html}}.
\newblock
\newblock
\shownote{(Accessed on 02/01/2023)}.


\bibitem[Guerriero et~al\mbox{.}(2019)]%
        {Guerriero2019}
\bibfield{author}{\bibinfo{person}{MIchele Guerriero}, \bibinfo{person}{Martin Garriga}, \bibinfo{person}{Damian~A. Tamburri}, {and} \bibinfo{person}{Fabio Palomba}.} \bibinfo{year}{2019}\natexlab{}.
\newblock \showarticletitle{Adoption, Support, and Challenges of Infrastructure-as-Code: Insights from Industry}. In \bibinfo{booktitle}{\emph{2019 IEEE International Conference on Software Maintenance and Evolution (ICSME)}}. \bibinfo{publisher}{IEEE}, \bibinfo{address}{New York City at 3 Park Ave}, \bibinfo{pages}{580--589}.
\newblock
\href{https://doi.org/10.1109/ICSME.2019.00092}{doi:\nolinkurl{10.1109/ICSME.2019.00092}}


\bibitem[Keller and Rexford(2010)]%
        {keller2010platform}
\bibfield{author}{\bibinfo{person}{Eric Keller} {and} \bibinfo{person}{Jennifer Rexford}.} \bibinfo{year}{2010}\natexlab{}.
\newblock \showarticletitle{The" Platform as a Service" Model for Networking.}
\newblock \bibinfo{journal}{\emph{INM/WREN}}  \bibinfo{volume}{10} (\bibinfo{year}{2010}), \bibinfo{pages}{95--108}.
\newblock


\bibitem[Krishna~Kaiser and Krishna~Kaiser(2018)]%
        {krishna2018managing}
\bibfield{author}{\bibinfo{person}{Abhinav Krishna~Kaiser} {and} \bibinfo{person}{Abhinav Krishna~Kaiser}.} \bibinfo{year}{2018}\natexlab{}.
\newblock \showarticletitle{Managing Configurations in a DevOps Project}.
\newblock \bibinfo{journal}{\emph{Reinventing ITIL{\textregistered} in the Age of DevOps: Innovative Techniques to Make Processes Agile and Relevant}} \bibinfo{volume}{0}, \bibinfo{number}{0} (\bibinfo{year}{2018}), \bibinfo{pages}{135--162}.
\newblock


\bibitem[Leite et~al\mbox{.}(2021)]%
        {leite2021organization}
\bibfield{author}{\bibinfo{person}{Leonardo Leite}, \bibinfo{person}{Gustavo Pinto}, \bibinfo{person}{Fabio Kon}, {and} \bibinfo{person}{Paulo Meirelles}.} \bibinfo{year}{2021}\natexlab{}.
\newblock \showarticletitle{The organization of software teams in the quest for continuous delivery: A grounded theory approach}.
\newblock \bibinfo{journal}{\emph{Information and Software Technology}}  \bibinfo{volume}{139} (\bibinfo{year}{2021}), \bibinfo{pages}{106672}.
\newblock


\bibitem[Luz et~al\mbox{.}(2019)]%
        {AdoptingDevOps2019}
\bibfield{author}{\bibinfo{person}{Welder~Pinheiro Luz}, \bibinfo{person}{Gustavo Pinto}, {and} \bibinfo{person}{Rodrigo Bonif{\'{a}}cio}.} \bibinfo{year}{2019}\natexlab{}.
\newblock \showarticletitle{Adopting DevOps in the real world: {A} theory, a model, and a case study}.
\newblock \bibinfo{journal}{\emph{J. Syst. Softw.}}  \bibinfo{volume}{157} (\bibinfo{year}{2019}), \bibinfo{pages}{0}.
\newblock
\href{https://doi.org/10.1016/j.jss.2019.07.083}{doi:\nolinkurl{10.1016/j.jss.2019.07.083}}


\bibitem[Mazyar(2018)]%
        {DevOpsTh9}
\bibfield{author}{\bibinfo{person}{Michael Mazyar}.} \bibinfo{year}{2018}\natexlab{}.
\newblock \bibinfo{title}{DevOps: The Ultimate Way to Break Down Silos - DevOps.com}.
\newblock \bibinfo{howpublished}{\url{https://devops.com/devops-the-ultimate-way-to-break-down-silos/}}.
\newblock
\newblock
\shownote{(Accessed on 01/27/2023)}.


\bibitem[Nazário et~al\mbox{.}(2025)]%
        {myGreyWork}
\bibfield{author}{\bibinfo{person}{Marcos Nazário}, \bibinfo{person}{Rodrigo Bonifácio}, \bibinfo{person}{Cleidson de Souza}, \bibinfo{person}{Fernando Kamei}, {and} \bibinfo{person}{Gustavo Pinto}.} \bibinfo{year}{2025}\natexlab{}.
\newblock \showarticletitle{Strategies to Mitigate Configuration Differences in Software Development: A Rapid Review of Grey Literature}.
\newblock \bibinfo{journal}{\emph{Journal of Software Engineering Research and Development}} \bibinfo{volume}{13}, \bibinfo{number}{1} (\bibinfo{date}{Feb.} \bibinfo{year}{2025}), \bibinfo{pages}{13:114 – 13:131}.
\newblock
\href{https://doi.org/10.5753/jserd.2025.4378}{doi:\nolinkurl{10.5753/jserd.2025.4378}}


\bibitem[Nazário et~al\mbox{.}(2023)]%
        {myCodes}
\bibfield{author}{\bibinfo{person}{Marcos Nazário}, \bibinfo{person}{Rodrigo Bonifácio}, {and} \bibinfo{person}{Gustavo Pinto}.} \bibinfo{year}{2023}\natexlab{}.
\newblock \bibinfo{booktitle}{\emph{Interview data}}.
\newblock
\href{https://doi.org/10.5281/zenodo.14767083}{doi:\nolinkurl{10.5281/zenodo.14767083}}


\bibitem[Nguyen(2006)]%
        {Nguyen2006}
\bibfield{author}{\bibinfo{person}{Tien~N. Nguyen}.} \bibinfo{year}{2006}\natexlab{}.
\newblock \showarticletitle{Model-oriented Configuration Management for Relational Database Applications}. In \bibinfo{booktitle}{\emph{The Sixth IEEE International Conference on Computer and Information Technology (CIT'06)}}. \bibinfo{publisher}{IEEE}, \bibinfo{address}{New York City at 3 Park Ave}, \bibinfo{pages}{194--194}.
\newblock
\href{https://doi.org/10.1109/CIT.2006.122}{doi:\nolinkurl{10.1109/CIT.2006.122}}


\bibitem[Overflow(2023)]%
        {StackOverflowSurvey23}
\bibfield{author}{\bibinfo{person}{Stack Overflow}.} \bibinfo{year}{2023}\natexlab{}.
\newblock \bibinfo{title}{Stack Overflow Developer Survey 2023}.
\newblock \bibinfo{howpublished}{\url{https://survey.stackoverflow.co/2023/\#developer-profile-experience}}.
\newblock
\newblock
\shownote{(Accessed on 07/12/2023)}.


\bibitem[Perera and Beck(2018)]%
        {perera2018continuous}
\bibfield{author}{\bibinfo{person}{Nadine Perera} {and} \bibinfo{person}{Thorsten Beck}.} \bibinfo{year}{2018}\natexlab{}.
\newblock \showarticletitle{Continuous delivery: Software deployment and configuration management for critical operations environments}. In \bibinfo{booktitle}{\emph{2018 SpaceOps Conference}}. \bibinfo{publisher}{American Institute of Aeronautics and Astronautics}, \bibinfo{address}{Marseille, France}, \bibinfo{pages}{2333}.
\newblock


\bibitem[Poth et~al\mbox{.}(2018)]%
        {poth2018deliver}
\bibfield{author}{\bibinfo{person}{Alexander Poth}, \bibinfo{person}{Mark Werner}, {and} \bibinfo{person}{Xinyan Lei}.} \bibinfo{year}{2018}\natexlab{}.
\newblock \showarticletitle{How to deliver faster with CI/CD integrated testing services?}. In \bibinfo{booktitle}{\emph{Systems, Software and Services Process Improvement: 25th European Conference, EuroSPI 2018, Bilbao, Spain, September 5-7, 2018, Proceedings 25}}. Springer, \bibinfo{publisher}{Springer, Cham}, \bibinfo{address}{none}, \bibinfo{pages}{401--409}.
\newblock


\bibitem[Rahman et~al\mbox{.}(2019)]%
        {RahmanHezaveh2019}
\bibfield{author}{\bibinfo{person}{Akond Rahman}, \bibinfo{person}{Rezvan Mahdavi-Hezaveh}, {and} \bibinfo{person}{Laurie Williams}.} \bibinfo{year}{2019}\natexlab{}.
\newblock \showarticletitle{A systematic mapping study of infrastructure as code research}.
\newblock \bibinfo{journal}{\emph{Information and Software Technology}}  \bibinfo{volume}{108} (\bibinfo{date}{apr} \bibinfo{year}{2019}), \bibinfo{pages}{65--77}.
\newblock
\href{https://doi.org/10.1016/j.infsof.2018.12.004}{doi:\nolinkurl{10.1016/j.infsof.2018.12.004}}


\bibitem[Rahman et~al\mbox{.}(2017)]%
        {rahman2017factors}
\bibfield{author}{\bibinfo{person}{Akond Rahman}, \bibinfo{person}{Asif Partho}, \bibinfo{person}{David Meder}, {and} \bibinfo{person}{Laurie Williams}.} \bibinfo{year}{2017}\natexlab{}.
\newblock \showarticletitle{Which factors influence practitioners' usage of build automation tools?}. In \bibinfo{booktitle}{\emph{2017 IEEE/ACM 3rd International Workshop on Rapid Continuous Software Engineering (RCoSE)}}. IEEE, \bibinfo{publisher}{IEEE}, \bibinfo{address}{New York City at 3 Park Ave}, \bibinfo{pages}{20--26}.
\newblock


\bibitem[Rahman et~al\mbox{.}(2018)]%
        {Rahman2018}
\bibfield{author}{\bibinfo{person}{Akond Rahman}, \bibinfo{person}{Asif Partho}, \bibinfo{person}{Patrick Morrison}, {and} \bibinfo{person}{Laurie Williams}.} \bibinfo{year}{2018}\natexlab{}.
\newblock \showarticletitle{What Questions Do Programmers Ask about Configuration as Code?}. In \bibinfo{booktitle}{\emph{2018 IEEE/ACM 4th International Workshop on Rapid Continuous Software Engineering (RCoSE)}}. \bibinfo{publisher}{IEEE}, \bibinfo{address}{New York City at 3 Park Ave}, \bibinfo{pages}{16--22}.
\newblock


\bibitem[Rahman2019 et~al\mbox{.}(2019)]%
        {Rahman2019}
\bibfield{author}{\bibinfo{person}{Akond Rahman2019}, \bibinfo{person}{Chris Parnin}, {and} \bibinfo{person}{Laurie Williams}.} \bibinfo{year}{2019}\natexlab{}.
\newblock \showarticletitle{The Seven Sins: Security Smells in Infrastructure as Code Scripts}. In \bibinfo{booktitle}{\emph{2019 IEEE/ACM 41st International Conference on Software Engineering (ICSE)}}. \bibinfo{publisher}{IEEE}, \bibinfo{address}{New York City at 3 Park Ave}, \bibinfo{pages}{164--175}.
\newblock
\href{https://doi.org/10.1109/ICSE.2019.00033}{doi:\nolinkurl{10.1109/ICSE.2019.00033}}


\bibitem[Shahin et~al\mbox{.}(2017)]%
        {shahin2017continuous}
\bibfield{author}{\bibinfo{person}{Mojtaba Shahin}, \bibinfo{person}{Muhammad~Ali Babar}, {and} \bibinfo{person}{Liming Zhu}.} \bibinfo{year}{2017}\natexlab{}.
\newblock \showarticletitle{Continuous integration, delivery and deployment: a systematic review on approaches, tools, challenges and practices}.
\newblock \bibinfo{journal}{\emph{IEEE access}}  \bibinfo{volume}{5} (\bibinfo{year}{2017}), \bibinfo{pages}{3909--3943}.
\newblock


\bibitem[Shahri et~al\mbox{.}(2007)]%
        {shahri2007software}
\bibfield{author}{\bibinfo{person}{Hamid~Haidarian Shahri}, \bibinfo{person}{James~A Hendler}, {and} \bibinfo{person}{Adam~A Porter}.} \bibinfo{year}{2007}\natexlab{}.
\newblock \showarticletitle{Software configuration management using ontologies}. In \bibinfo{booktitle}{\emph{3rd International Workshop on Semantic Web Enabled Software Engineering (SWESE 2007), Innsubruk, Austria}}. Citeseer, \bibinfo{publisher}{UMIACS}, \bibinfo{address}{8125 Paint Branch Dr, College Park}, \bibinfo{pages}{1}.
\newblock


\bibitem[Strandberg(2019)]%
        {Strandberg2019}
\bibfield{author}{\bibinfo{person}{Per~Erik Strandberg}.} \bibinfo{year}{2019}\natexlab{}.
\newblock \showarticletitle{Ethical Interviews in Software Engineering}. In \bibinfo{booktitle}{\emph{2019 ACM/IEEE International Symposium on Empirical Software Engineering and Measurement (ESEM)}}. \bibinfo{publisher}{IEEE}, \bibinfo{address}{New York City at 3 Park Ave}, \bibinfo{pages}{1--11}.
\newblock
\href{https://doi.org/10.1109/ESEM.2019.8870192}{doi:\nolinkurl{10.1109/ESEM.2019.8870192}}


\bibitem[Tichy(1995)]%
        {tichyConfiguration1995}
\bibfield{author}{\bibinfo{person}{Walter~F Tichy}.} \bibinfo{year}{1995}\natexlab{}.
\newblock \bibinfo{booktitle}{\emph{Configuration management}}.
\newblock \bibinfo{publisher}{John Wiley \& Sons, Inc.}, \bibinfo{address}{111 River St, Hoboken, New Jersey}.
\newblock


\bibitem[Zeng et~al\mbox{.}(2014)]%
        {zeng2014managing}
\bibfield{author}{\bibinfo{person}{Sai Zeng}, \bibinfo{person}{Constantin Adam}, \bibinfo{person}{Fred Wu}, \bibinfo{person}{Shang Guo}, \bibinfo{person}{Yaoping Ruan}, \bibinfo{person}{Cashchakanithara Venugopal}, {and} \bibinfo{person}{Rajeev Puri}.} \bibinfo{year}{2014}\natexlab{}.
\newblock \showarticletitle{Managing risk in multi-node automation of endpoint management}. In \bibinfo{booktitle}{\emph{2014 IEEE Network Operations and Management Symposium (NOMS)}}. IEEE, \bibinfo{publisher}{IEEE}, \bibinfo{address}{New York City at 3 Park Ave}, \bibinfo{pages}{1--6}.
\newblock


\end{thebibliography}


\end{document}